\documentclass[10pt,conference]{IEEEtran}

\IEEEoverridecommandlockouts

\usepackage{amsmath,amssymb,amsfonts}
\usepackage{algorithmic}
\usepackage{graphicx}
\usepackage{textcomp}
\usepackage{xcolor}
\usepackage{ragged2e}
\usepackage{balance}
\usepackage{colortbl}
\usepackage{multirow}
\usepackage{balance}
\usepackage{color}
\usepackage{float}
\usepackage{fancybox}
\usepackage{placeins}
\usepackage{graphicx}

\usepackage{CJKutf8}
\usepackage{indentfirst}
\usepackage{tcolorbox}
\tcbuselibrary{skins, breakable, theorems}

\usepackage[figuresright]{rotating}
\usepackage{rotfloat}
\usepackage{booktabs}

\usepackage[colorlinks,
            linkcolor=blue,
            anchorcolor=blue,
            citecolor=blue]{hyperref}

\begin{document}

\title{Understanding Resolution of Multi-Language Bugs: \\An Empirical Study on Apache Projects}


\author{
    \IEEEauthorblockN{Zengyang Li\IEEEauthorrefmark{2}\thanks{This work was funded by the Natural Science Foundation of Hubei Province of China under Grant No. 2021CFB577, the National Natural Science Foundation of China under Grant Nos. 62176099 and 62172311, and the Knowledge Innovation Program of Wuhan-Shuguang Project under Grant No. 2022010801020280.}, Wenshuo Wang\IEEEauthorrefmark{2}, Sicheng Wang\IEEEauthorrefmark{2}, Peng Liang\IEEEauthorrefmark{3}\IEEEauthorrefmark{1}, Ran Mo\IEEEauthorrefmark{2}}
    \IEEEauthorblockA{
    \IEEEauthorrefmark{2}School of Computer Science \& Hubei Provincial Key Laboratory of Artificial Intelligence and Smart Learning, \\Central China Normal University, Wuhan, China \\
    \IEEEauthorrefmark{3}School of Computer Science, Wuhan University, Wuhan, China\\
    \{zengyangli, moran\}@ccnu.edu.cn, \{scwang1998, wenshuowang\}@mails.ccnu.edu.cn, liangp@whu.edu.cn}
}


\IEEEoverridecommandlockouts
\IEEEpubid{\makebox[\columnwidth]{978-1-6654-5223-6/23/\$31.00~\copyright2023 IEEE\hfill} \hspace{\columnsep}\makebox[\columnwidth]{ }}
\maketitle 
\IEEEpubidadjcol

\begin{abstract}
\textit{Background}: In modern software systems, more and more systems are written in multiple programming languages (PLs). There is no comprehensive investigation on the phenomenon of multi-programming-language (MPL) bugs, which resolution involves source files written in multiple PLs.
\textit{Aim}: This work investigated the characteristics of bug resolution in MPL software systems and explored the reasons why bug resolution involves multiple PLs.
\textit{Method}: We conducted an empirical study on 54 MPL projects selected from 655 Apache OSS projects, of which 66,932 bugs were analyzed. 
\textit{Results}: (1) the percentage of MPL bugs (MPLBs) in the selected projects ranges from 0.17\% to 42.26\%, and the percentage of MPLBs for all projects as a whole is 10.01\%; (2)  95.0\% and 4.5\% of all the MPLBs involve source files written in 2 and 3 PLs, respectively; (3) the change complexity resolution characteristics of MPLBs tend to be higher than those of single-programming-language bugs (SPLBs); (4) the open time for MPLBs is 19.52\% to 529.57\% significantly longer than SPLBs regarding 9 PL combinations; (5) the reopen rate of bugs involving the PL combination of JavaScript and Python reaches 20.66\%; (6) we found 6 causes why the bug resolution involves multiple PLs and identified 5 cross-language calling mechanisms.
\textit{Conclusion}: MPLBs are related to increased development difficulty.  

\end{abstract}




\section{Introduction}
\label{chap:intro}

In modern software systems, more and more systems are developed in multiple programming languages (PLs)  \cite{kontogiannis2006comprehension, mayer2015empirical, kochhar2016large, abidi2021multi, li2022exploring}. We call such systems multi-programming-language (MPL) software systems, which can take advantage of each PL and reuse existing code and libraries to meet various quality requirements and to improve software development efficiency \cite{kochhar2016large, abidi2021multi, li2022exploring, ray2014large, abidi2019behind, grichi2020impactJNI}. During the development of an MPL system, a bug may be fixed by one or more bug-fixing commits in which source files written in multiple PLs are involved, and such a bug is called as an MPL bug (MPLB). In contrast, if a bug is fixed by one or more commits in which source files in the same PL are involved, then such a bug is called as a single-PL bug (SPLB). 

MPLBs may impact a software system at the architecture level, given that the resolution of an MPLB may involve comprehending, modifying, debugging, and testing inter-language communications and inter-component interactions.
To our knowledge, there are only two works closely related to MPLBs and their impact on software development. Li et al. explored the impact of MPL commits (MPLCs) on development difficulty and software quality in 18 Apache MPL OSS projects \cite{li2022exploring}. They found that the issues (including new features, bugs, improvements, etc.) fixed in MPLCs take longer time to be fixed than issues fixed in non-MPLCs, and there is no significant differences between MPLCs and non-MPLCs with respect to the impact on issue reopen in most projects \cite{li2022exploring}. 
However, MPLBs were not the core research objects in that work, and thus no in-depth or comprehensive investigation on MPLBs was performed.
In the other work, Li et al. looked into 1,497 bugs with 406 MPLBs included in 3 MPL deep learning frameworks  \cite{li2023understanding}. Although some bug resolution characteristics of MPLBs were investigated in that work, the numbers of MPLBs and projects are relatively small, and PL combinations involved and in-depth analysis of causes for involving multiple PLs in MPLB resolution were not considered.

To understand the resolution of MPLBs, we comprehensively investigated the bug resolution characteristics of Apache OSS projects in an MPL context and explores the reasons why bug resolution involves multiple PLs. We conducted a large-scale empirical study on 54 non-trivial MPL projects selected from 655 Apache OSS projects, of which 66932 bugs were analyzed, including 6700 MPLBs. 

The \textbf{main contributions} of this paper are: (1) This work is the first attempt to comprehensively explore the phenomenon of MPLBs in real-world projects.
(2) This work presents a large-scale empirical study on 54 non-trivial Apache MPL OSS projects.
(3) This work investigated resolution characteristics of MPLBs with 13 PL combinations in terms of change complexity, open time, and reopen.
(4) This work explored the reasons why bug resolution involving multiple PLs and the cross-language calling mechanisms used.


\section{Related Work}\label{RelatedWork}
\label{chap:relat}
The related work centers around the phenomenon of MPL software systems and bug resolution characteristics.

\subsection{Phenomenon of MPL Software Systems}\label{chap:relat_phenomenon}


Several studies investigated the practice of PL selection in MPL systems. In 2014, Tomassetti et al. found that 96\% of projects used at least two PLs, usually a combination of PLs, such as C/C++, Makefile and HTML, CSS, JavaScript, and over 50\% of projects used at least two general purpose languages (GPLs) \cite{tomassetti2014empirical}. In 2015, Mayer et al. identified three language ecosystems that are associated with the GPL: (a) Shell and Make, (b) XML, and (c) HTML and CSS \cite{mayer2015empirical}. Similar to the study by Bissyandé et al. \cite{bissyande2013popularity}, in 2021, Li et al. studied the PL selection and usage of multilingual OSS projects on GitHub spanning 10 years (2010-2019) \cite{li2021understanding}, and they found that there exists a veriﬁable correlation between software domains and sets of mainstream PLs. Studies also show that the increasing adoption of multiple PLs is to take advantage of different PLs and meet development requirements \cite{meyerovich2013empirical, abidi2019behind}. Although in our work we also explored PLs used in Apache MPL OSS projects, we paid more attention to resolution characteristics of bugs involving different PL combinations. 

In 2017, Mayer et al. launched a survey of 139 professional software developers \cite{mayer2017multi}. Respondents saw benefits of multi-language development for the motivation of developers and the translation of requirements, but the existence cross-language linking is in many cases preventing them from making necessary changes to the source code.
The presence of cross-language links has prompted researchers to explore ways to recognize and mitigate the associated threats of multiple PLs. 

Some studies investigated cross-language relations between specific PL combinations, while others attempted to develop a taxonomy of cross-language links. Cross-language analysis between Java and C (i.e., using Java Native Interface (JNI)) has been studied relatively extensively. In 2020, Grichi et al. proposed two approaches for multi-language dependency analysis of JNI systems \cite{grichi2020impactJNI}: one approach is to use JNI rules to statically parse a software system and generate a model that incorporates all of its constituent parts; the other approach is to analyze the MPL source code as it changes together, at the commit level. There are some similar JNI studies \cite{lee2020broadening, shen2021cross}. There was also cross-language analysis between Python and C \cite{monat2021multilanguage}, which uses multi-language static syntax analysis for Python and C to parse Python programs with native C modules. 
In 2017, Mayer analyzed 22 open-source frameworks and classified the cross-language linking mechanisms among GPLs and DSLs, resulting in a taxonomy of cross-language linking mechanisms \cite{mayer2017taxonomy}. In 2022, Li et al. defined four language interface categories to represent the multilingual calling mechanisms between languages \cite{li2022vulnerability}. In contrast, our work also defines cross-language calling mechanisms, and gives real-world examples to help stakeholders better understand these mechanisms.
\subsection{Bug Resolution Characteristics}
There are a number of studies that are closed to bug resolution characteristics.
In 2011, Bhattacharya et al. analyzed four OSS projects written in C and C++ \cite{bhattacharya2011assessing}. They used the number of lines of code changed during bug resolution to measure maintainability and found that using C++ improved software quality and reduced maintenance effort, and the code base is shifting from C to C++. 
In 2019, Zhang et al. examined three bug-resolution characteristics of 10 GPLs, namely SLOC changed, files touched, and bug-resolution time \cite{zhang2019study}. Their results revealed that bug-resolution time was shorter for projects written in Java compared to other PLs. In contrast, for projects written in Ruby, the bug-resolution time was longer. Their study focused on bug resolution from a single-language perspective, our work focuses on the resolution characteristics of MPLBs, exploring the potential impact of MPLBs on software development through six bug resolution characteristics. 
In 2020, Li et al. investigated whether bug severity is in line with code change complexity of bug resolution, which was measured by four bug resolution characteristics for Java projects: the number of modified lines of code, the number of modified source files, the number of modified packages, and the entropy of the change process \cite{li2020bug}. They found that there was not constantly consistent between bug severity and code change complexity of bug resolution. In 2023, Li et al. used open time, three change complexity measures of pull requests, and two communication complexity measures in bug resolution, to explore the impact of bugs (including MPLBs) on development in three deep learning frameworks \cite{li2023understanding}.

Building on the aforementioned studies on bug resolution characteristics, our study incorporates the bug resolution characteristics of 4 code change complexity measures and open time. We specifically include the resolution characteristic of bug reopen due to its key impact on software maintainability.

\section{Study Design}
\label{chap:case}
We describe the empirical study designed and reported following the guidelines proposed by Runeson and H{\"o}st \cite{RuHo2009}.

\subsection{Objective and Research Questions}\label{DesignRQ}

The goal of this study described using the Goal-Question-Metric (GQM) approach \cite{Ba1992} is: to analyze bugs as well as their corresponding modified source files for the purpose of exploration with respect to the state of MPLBs as well as their resolution characteristics, from the point of view of software developers in the context of MPL OSS development.

Based on the above mentioned goal, we formulated 5 research questions (RQs) as shown in TABLE \ref{table-RQs}. RQ1 and RQ2 investigate the state of MPLBs. RQ3 and RQ4 center around the resolution characteristics of MPLBs and explores the possible impact of MPLBs on development difficulty. RQ5 explores possible reasons for why the resolution of MPLBs involves multiple PLs.

\begin{table*}[]
\caption{Research Questions and Their Rationales.}
\scriptsize
\centering
\begin{tabular}{p{0.06\columnwidth} p{0.6\columnwidth} p{1.2\columnwidth}}
\toprule
\textbf{\#} & \multicolumn{1}{c}{\textbf{RQ}} & \multicolumn{1}{c}{\textbf{Rationale}} \\ \midrule
RQ1 & What is the percentage of resolved MPLBs over all resolved bugs in an MPL system? & With this RQ, we investigate the frequency of MPLBs occurring in MPL systems to have a basic understanding of the status of MPLBs in MPL systems. \\
RQ2 & How many PLs are used for MPLB resolution? & This RQ examines the number of PLs involved in the MPLB resolution and the distribution of MPLBs involving different numbers of PLs. This allows us to understand the tendency of the use of multiple PLs in MPLB resolution. \\
RQ3 & What are the resolution characteristics of MPLBs? & To evaluate the possible impact of MPLBs on development difficulty, we explore the resolution characteristics of MPLBs in this RQ by analyzing the bug resolution process data. The resolution characteristics that we examine include change complexity, open time, and reopen. \\
RQ4 & What are the differences on the resolution characteristics of MPLBs with different PL combinations in MPL software systems? & For MPL systems, understanding the similarities and differences in the bug resolution of MPLBs involving different PL combinations can facilitate release planning and task assignment. \\
RQ5 & Why do the resolution of the MPLBs in the selected projects involve multiple PLs? & Resolution of MPLBs requires modifying source files in multiple PLs. With this RQ, we explore the rationales behind the co-changes of source files in different PLs and mine the possible logic links \cite{tomassetti2013classification} between such source files. The results will deepen our understanding on the resolution of MPLBs. \\ \bottomrule
\end{tabular}
\label{table-RQs}
\end{table*}


\subsection{Project Selection}\label{CaseSelection}
In this study, we only investigated Apache MPL OSS projects, since the links between issues and corresponding commits tend to be well recorded in the commit messages of those projects. For selecting each project included in our study, we applied the following inclusion criteria:

\begin{table}[]
\caption{Programming languages examined.}
\scriptsize
\centering
\begin{tabular}{cccccc}
\toprule
\# & PL           & \# & PL          & \# & PL         \\ \midrule
1  & C/C++        & 7  & Haskell     & 13 & PHP        \\ 
2  & C\#          & 8  & Java        & 14 & Python     \\ 
3  & Clojure      & 9  & JavaScript  & 15 & Ruby       \\ 
4  & CoffeeScript & 10 & Kotlin      & 16 & Scala      \\ 
5  & Erlang       & 11 & Objective-C & 17 & Swift      \\ 
6  & Go           & 12 & Perl        & 18 & TypeScript \\ 
\bottomrule
\end{tabular}
\label{table:ProgrammingLanguages}
\end{table}

\begin{itemize}
\item[\textbf{C1:}] The project uses JIRA \cite{JIRA} as its issue tracking system. 
This criterion is set to ensure that the issues from different projects have the same format so that issues can be handled in the same way. It is convenient to export issue data through the REST API provided by JIRA \cite{JIRAAPI}. 
\item[\textbf{C2:}] The project has more than 1000 issue records in JIRA. This criterion was set to ensure the dataset was big enough to be statistically analyzed.
\item[\textbf{C3:}] At least 2 out of the 18 PLs listed in TABLE \ref{table:ProgrammingLanguages} are used in the project. All the 18 listed PLs are mainstream general-purpose PLs, which are adopted from \cite{li2022exploring}. The percentage of each PL is greater than 5\%, and the percentage of the main PL does not exceed 90\%.  
\end{itemize}

\subsection{Data Collection}\label{DataCollection}
\subsubsection{Data Items to be Collected}
\label{dataitems}

\begin{table*}[]
\caption{Data items to be collected for each bug.}
\scriptsize
 \centering

\begin{tabular}{p{0.06\columnwidth} p{0.25\columnwidth} p{1.22\columnwidth} p{0.25\columnwidth} }
\toprule
\textbf{\#} & \textbf{Name} & \textbf{Description}    & \textbf{RQ}     \\ \midrule
D1          & BugID  & The unique ID of the bug in the Apache projects.
  & RQ1-RQ5     \\ 
D2          & IsMPLB & Whether the bug is an MPLB.  & RQ1-RQ5  \\ 
D3          & PL & The name of the PLs involved in the resolution of the bug. & RQ2, RQ4, RQ5  \\ 
D4          & PLNo & The number of PLs involved in the resolution of the bug.   & RQ2, RQ4, RQ5  \\ 
D5          & LOCM          & The number of lines of source code modified in the bug. & RQ3-RQ4  \\ 
D6          & NOFM          & The number of source files modified in the bug.  & RQ3-RQ4  \\ 
D7          & NODM          & The number of directories modified in the bug.  & RQ3-RQ4   \\
D8          & OT            & The time from the creation of a bug report to the final resolution of the bug.         & RQ3-RQ4        \\
D9          & Entropy       & The normalized entropy of the modified source files for  fixing the bug during the last 60 days.                 & RQ3-RQ4     \\ 
D10         & Reopen        & Whether the bug was reopened.         & RQ3-RQ4        \\
D11         & Summary       &   A brief one-line summary of the bug.     & RQ5   \\
D12         & Description   &   A detailed description on JIRA of the bug. & RQ5   \\
D13         & Commit   &   The code changes and commit messages during bug resolution.     & RQ5   \\

\bottomrule
\end{tabular}
\label{table:DataItemsForBug}
\end{table*}

To answer the RQs, we took a bug as the unit of analysis and the data items to be collected are listed in Table \ref{table:DataItemsForBug}. All the data items to be collected except for D2 and D9 are straightforward, thus we only explain D2 (IsMPLB) and D9 (Entropy) data items in detail. 

First, we explain data item D2. A bug is fixed by one or more commits. There are two types of commits: (1) MPL commit (MPLC), in which source files in multiple PLs are modified, and (2) Single-PL commit (SPLC), in which source files in the same single PL are modified.
To determine a bug is an MPLB, it should satisfy one of the following conditions: 1) the bug is resolved in one or more commits, in which at least one commit is an MPLC; 2) the bug is resolved in multiple SPLCs involving different PLs. 

Then, we explain the definition of the entropy of the modified source files in a bug resolution (i.e., D9) \cite{Ha2009}. 
Suppose that the modified source files of commit $c$ are  $\{f_1,f_2,\cdots,f_n\}$, and file $f_i\left(1\leq i\leq n\right)$ was modified in $t_i$ commits during a period of time  before the commit. Let $p_i = t_i/\sum_{i=1}^nt_i .$
\noindent Then, the entropy $H(m) = -\sum_{i=1}^mp_ilog_2 p_i .$
\noindent $m$ indicates the number of files modified in the commit(s) to fix the bug. 
Since the number of modified source files differs between different periods, we need to normalize the entropy to be comparable. Given that $H(m)$ achieves the maximum of $log_2 m$ when $p_i=1/m$ $(1\leq i\leq m)$, the normalized entropy
\begin{equation}
\widetilde{H}(m)=
\begin{cases}
H(m)/log_2 m & m>1 ,\\
0 &m=1 .
\end{cases}
\end{equation}
In this study, the period is set to 60 days (including the day when commit $c$ happened), which is chosen according to \cite{li2020bug}.

\subsubsection{Data Collection Procedure}\label{datacollectionprocedure}
The data collection procedure for each selected project consists of six steps. 

\noindent \textbf{Step 1}: Export bug reports. We used JIRA REST API to export all the bugs of the project. 

\noindent \textbf{Step 2}: Store the exported bug reports from Step 1 in MySQL.

\noindent \textbf{Step 3}: Clone the Git repository of the project from GitHub. 

\noindent \textbf{Step 4}: Extract the commit records from the Git repository. Extract the commit records from the project's Git repository into a text file, which is formatted for further parsing. In this step, we only exported the commit records of the master branch and the commit records that were merged into the master branch. The commit records corresponding to the MERGE operations were excluded, because the commit record corresponding to the MERGE operation is duplicated with the merged commit records.

\noindent \textbf{Step 5}: Match each bug report with corresponding commit record(s). If a commit is to resolve a certain bug, the committer often explicitly mention the bug ID in the commit message. Thus, a bug can be matched with corresponding commit record(s) through the bug ID.

\noindent \textbf{Step 6}: Calculate data items listed in TABLE \ref{table:DataItemsForBug} for each bug. 


\subsection{Data Analysis}

The answers to RQ1 and RQ2 can be obtained by descriptive statistics. To answer RQ3, the resolution characteristics of MPLBs were compared with those of SPLBs in order to observe the impact of the introduction of multiple PLs on bug resolution. 
To answer RQ4, we first present the resolution characteristics of MPLBs with different PL combinations, and then compared the resolution characteristics of MPLBs involving different PL combinations with those of SPLBs involving a single PL from the corresponding PL combinations. 
To do the comparisons for RQ3 and RQ4, we performed the Mann-Whitney U test to check whether two sample groups of data are significantly different \cite{Fi2013}. We also conducted the Chi-squared test, and the two variables of a Chi-squared test are: whether a fixed bug was reopened or not and whether it is an MPLB or not. 
The test is significant at \textit{p-value} \textless 0.05, which means that the two variables are connected. 

To answer RQ5, 
we needed to obtain the category of the resolution of each MPLB and the cross-language calling mechanism used in the source files modified. According to \cite{li2022exploring}, the proportion of commits involving 3 or more PLs is rather small, thus we divided all the MPLBs into 3 parts, i.e., MPLBs which resolution involves 2 PLs, 3 PLs, and 4 and more PLs. 
If the number of any of the 3 parts of the MPLBs is not greater than 500, we conducted manual analysis on all the MPLBs; otherwise, we manually analyzed a sample of the MPLBs, and the sample size was determined by taking 99\% of confidence level and 5\% of margin of error \cite{israel1992dss}. 
The first three authors worked together to label the causes of bug resolution as tags, and each tag represents a category. The process of manual analysis on each MPLB is described as follows: 
 
\noindent \textbf{Step 1}: Check the bug summary. We first read the bug summary to get a rough idea on what the MPLB is about. 

\noindent \textbf{Step 2}: Check the bug-fixing commit(s). We carefully read the commit messages and code changes of the bug-fixing commit(s) of the MPLB. If the tag can be determined, then the analysis of this MPLB is finished; otherwise, go to the next step.

\noindent \textbf{Step 3}: Check the description of the MPLB. We synthesized the description with the knowledge obtained in Step 2. If the tag can be determined, then the analysis of this bug is finished; otherwise, go to the next step.

\noindent \textbf{Step 4}: Repeat Steps 2 and 3 till the tag can be determined. 

Then, we describe the classification process of all the sampled MPLBs as follows. 

\noindent \textbf{Step S1}: Construct a preliminary set of tags. the second and third authors independently labeled 100 MPLBs out of the sampled MPLBs. Then the first author joined them discussed to address any inconsistencies, and to construct a preliminary set of tags for the MPLBs. 

\noindent \textbf{Step S2}: Conduct a pilot MPLB labelling. The second and third authors labeled 100 MPLBs with the preliminary set of tags independently. In this step, new MPLB tags might arise, existing tags might be merged and removed. If there was any disagreement on MPLB labelling, they discussed with the first author to get a consensus. 
Fleiss Kappa was used to measure the consistency between MPLB labelling results of the two authors \cite{fleiss1973equivalence}. If the Kappa value is less than 0.75, the two authors needed to discuss to resolve disagreements, and randomly selected another 100 MPLBs for another round of labelling. This iterative labelling process would stop when the Kappa value exceeds 0.75, indicating substantial agreement. 

\noindent \textbf{Step S3}: Classify the remaining MPLBs. The second author labelled the remaining MPLBs with the updated set of tags.

After labelling the sampled MPLBs with causes for involving multiple PLs, we further identified the cross-language calling mechanisms used by the modified source files in different PLs during bug resolution. This may help further understand the underlying reasons for involving multiple PLs in bug resolution \cite{mayer2017taxonomy, li2022vulnerability}. Since the used cross-language calling mechanism in the MPLB resolution is not ambiguous, it could be identified by analyzing the source files. 

\section{Study Results}
\label{chap:study}

Based on the criteria defined in Section \ref{CaseSelection}, 655 Apache projects were obtained by criterion C1, 184 projects were left by C2, and 54 projects were finally included for data extraction and analysis according to C3.
The basic information of the 54 projects is shown in an online table \cite{ProjectInfo}. We then collected the data items described in Table \ref{table:DataItemsForBug} from the 54 projects, and the data were collected during July to August, 2022.

\subsection{Proportion of MPLBs in the Selected Projects (RQ1)}

\begin{figure*}
\centering
\centerline{\includegraphics[width=6.4in]{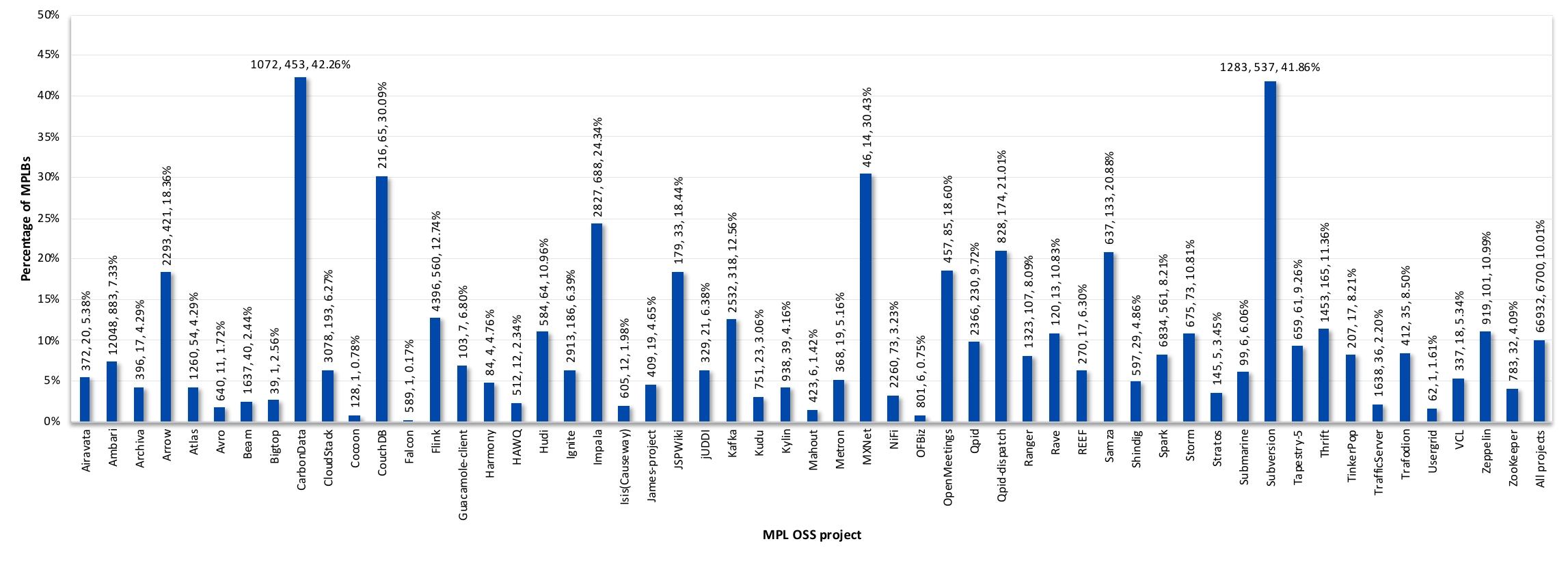}}
\caption{Proportion of MPLBs in each selected project (RQ1).}
\label{fig-RQ1}
\end{figure*}

Fig. \ref{fig-RQ1} shows the percentage of the number of MPLBs over the number of bugs explicitly associated with commits in each project. In this figure, for each project, the three numbers denote the number of bugs that are explicitly associated with commits, the number of MPLBs, and the percentage of the former over the latter. There are 6,700 MPLBs over 66,932 bugs in all the selected Apache projects, and the percentage of MPLBs is 10.01\% when considering the bugs of all projects as a whole. The percentage of MPLBs in each project varies from 0.17\% (project \textit{Falcon}) to 42.26\% (project \textit{CarbonData}).

\subsection{Number of PLs Used in the Resolution of MPLBs (RQ2)}


We explored the distribution of the 18 PLs used in the resolution of MPLBs, and the results are presented in an online table \cite{RQ2Table}. Taking all the selected projects as a whole, 95.07\% and 4.51\% of the MPLBs involve source files in 2 and 3 PLs, respectively; this means that the vast majority of MPLBs involve source files written in only 2 PLs, and it is uncommon to modify source files in 4 or more PLs in MPLB resolution.

\subsection{Resolution Characteristics of MPLBs (RQ3)}

\begin{table}[]
\caption{Comparison of resolution characteristics between MPLBs and SPLBs (RQ3).}
\scriptsize
\centering
\begin{tabular}{ccccc}
\toprule
                 & \textbf{MPLB} & \textbf{SPLB} & \textbf{\%Diff} & \textbf{\textit{p-value}} \\
                 \midrule
\textbf{LOCM}    & 110.00           & 30.00            & 266.67\%        & \textless 0.001  \\
\textbf{NOFM}    & 5.00             & 2.00             & 150.00\%        & \textless 0.001  \\
\textbf{NODM}    & 3.00             & 2.00             & 50.00\%         & \textless 0.001  \\
\textbf{OT}      & 9.46         & 4.27         & 121.55\%        & \textless 0.001  \\
\textbf{Entropy} & 0.92        & 0.84        & 9.52\%          & \textless 0.001  \\
\textbf{Reopen Rate}  & 0.065       & 0.056       & 16.07\%         & 0.004   \\
\bottomrule
\end{tabular}
\label{table-compare-MPLB-SPLB}
\end{table}

To better understand the resolution characteristics of MPLBs, we conducted a comparison of the resolution characteristics for the 6,700 MPLBs and 60,232 SPLBs, which is summarized in Table \ref{table-compare-MPLB-SPLB}. Columns MPLB and SPLB present the medians of LOCM, NOFM, NODM, OT (in days), Entropy, and Reopen Rate in the corresponding rows of the MPLBs and SPLBs, respectively.
Column \%Diff presents the percentage of difference between \emph{MPLB} and \emph{SPLB}, i.e.,
$\%Diff=(MPLB-SPLB)/SPLB\times{100\%} .$
Column \emph{p-value} reports the results of the Mann-Whitney U tests on the first five resolution characteristic between MPLBs and SPLBs. For Reopen Rate, \emph{p-value} reports the result of the Chi-squared test, and the test results indicate that whether a bug has been reopened or not is related to whether the bug resolution involves multiple PLs.
As shown in Table \ref{table-compare-MPLB-SPLB}, the LOCM, NOFM, NODM, OT, and Entropy of MPLB resolution are significantly larger than those of SPLB resolution. In addition, there is a significant connection between bug reopen and resolution involving multiple PLs.


\subsection{Resolution Characteristics of MPLBs with Different PL Combinations (RQ4)}

We grouped all MPLBs by different PL combinations, and obtained 104 PL combinations in total. To ensure that the dataset for each PL combination is big enough for statistical analysis, only the PL combinations with more than 50 MPLBs are considered. The resulting 13 PL combinations with more than 50 MPLBs are shown in Table \ref{table-RQ4-PLsCombination}, where \#MPLB denotes the number of MPLBs whose bug-fixing commits involve the corresponding PL combination. 

\begin{table}[]
\caption{The number of MPLBs of PL combinations involved in the bug-fixing commits of MPLBs in the selected projects (RQ4).}
\scriptsize
\centering

\begin{tabular}{cccccc}
\toprule
\textbf{\#} & \textbf{PL Combination}  & \textbf{\#MPLB} & \textbf{\#} & \textbf{PL Combination}  & \textbf{\#MPLB} \\ \midrule
1  & Java,Scala      & 1782   & 8  & JavaScript,Python      & 121    \\
2  & C/C++,Python    & 1635   & 9  & C/C++,Java,Python      & 86     \\
3  & Java,Python     & 903    & 10 & Erlang,JavaScript      & 63     \\
4  & Java,JavaScript & 799    & 11 & JavaScript,PHP         & 62     \\
5  & Python,Scala    & 266    & 12 & Clojure,Java           & 57     \\
6  & C/C++,Java      & 246    & 13 & Java,JavaScript,Python & 52     \\
7  & C\#,Java        & 122    &    & Total                  & 6194   \\ \bottomrule
\end{tabular}
\label{table-RQ4-PLsCombination}
\end{table}

\begin{figure}
\centerline{\includegraphics[width=3.0in]{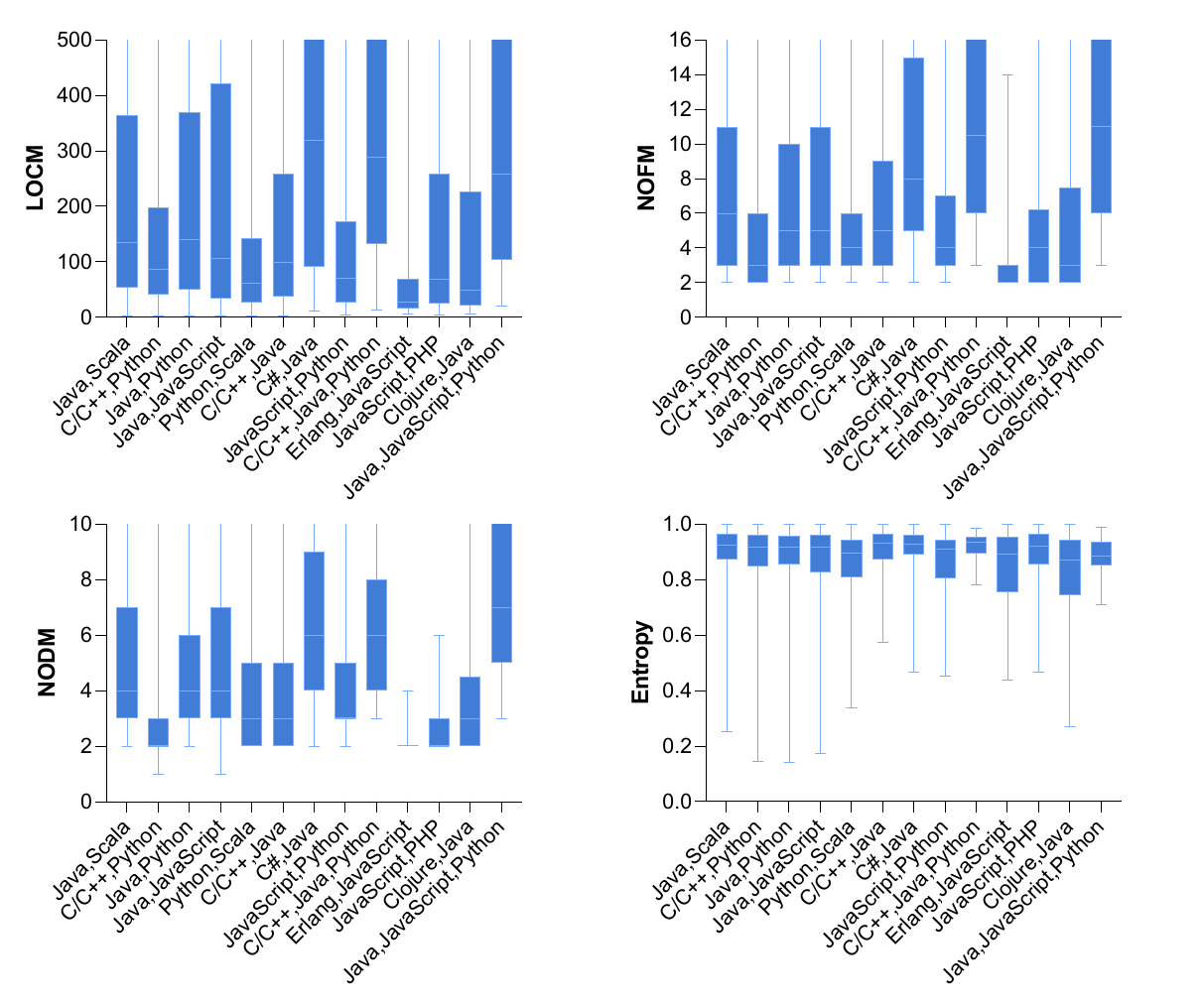}}
\caption{Distributions of change complexity measures of bug-fixing commits of MPLBs with different PL combinations (RQ4).}
\label{fig-Bug-ChangeComplexity}
\end{figure}

\begin{figure}
\centerline{\includegraphics[width=3.2in]{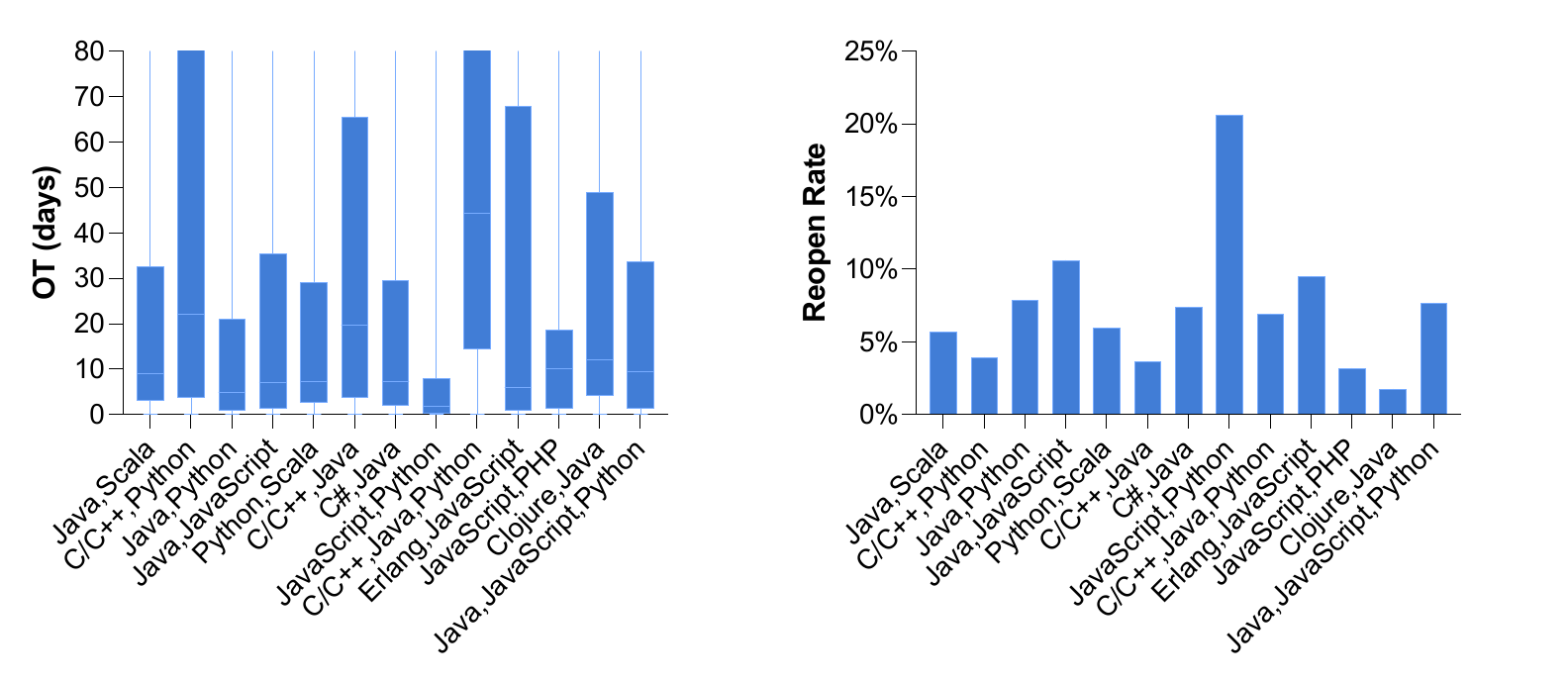}}
\caption{Distribution of the OT and Reopen Rate of MPLBs with different PL combinations (RQ4).}
\label{fig-Bug-Reopen-OT}
\end{figure}

Based on the 13 PL combinations shown in Table \ref{table-RQ4-PLsCombination}, we analyzed bug resolution characteristics for the bugs of each PL combination, including the four measures for change complexity of bug-fixing commits (i.e., LOCM, NOFM, NODM, and Entropy), OT, and Reopen Rate of bugs. The distributions of the four change complexity measures are shown in Fig. \ref{fig-Bug-ChangeComplexity}, and the distributions of Reopen Rate and OT are shown in Fig. \ref{fig-Bug-Reopen-OT}. In these two figures, the horizontal coordinates indicate the 13 PL combinations in Table \ref{table-RQ4-PLsCombination} and the vertical coordinates indicate the four change complexity measures, the values of Reopen Rate and OT, respectively.

In the rest of this section, we first present the results on the resolution characteristics of MPLBs with different PL combinations, and then compare the resolution characteristics of MPLBs and those of SPLBs with different PL combinations.

\subsubsection{Resolution characteristics of MPLBs}


From Fig. \ref{fig-Bug-ChangeComplexity} and Fig. \ref{fig-Bug-Reopen-OT}, we can see that the distributions of LOCM, NOFM, NODM, and OT of MPLBs differ apparently between different PL combinations involved.
The Reopen Rate of MPLBs are also obviously different between different PL combinations involved.
However, Fig. \ref{fig-Bug-ChangeComplexity} 
shows relatively large Entropy medians of the MPLBs involving all the 13 PL combinations, given that all the Entropy medians are either near 0.900 or greater than 0.900.

In Fig. \ref{fig-Bug-ChangeComplexity} and Fig. \ref{fig-Bug-Reopen-OT}, it can be observed that PL combinations involving Java (Java, Scala; Java, Python; Java, JavaScript; C\#, Java) have a larger median LOCM, NOFM, and NODM than other PL combinations, and tend to have a lower median OT.
PL combinations involving C/C++ (C/C++, Python; C/C++, Java) demonstrate a greater median OT than other PL combinations, and tend to have a smaller median LOCM, NOFM, and NODM. 
The combinations of three PLs (Java, JavaScript, Python; C/C++, Java, Python) show a relatively large median of LOCM, NOFM, and NODM. 


\subsubsection{Comparison of Resolution Characteristics between MPLBs and SPLBs}
Next, we investigated the differences on the resolution characteristics between MPLBs and SPLBs.
In Table \ref{table-MPLBs-SPLBs}, MedianSPLB denotes the maximum median of the corresponding resolution characteristic of the SPLBs involving each single PL in the corresponding PL combination, and MedianMPLB denotes the median of the corresponding resolution characteristic of the MPLBs involving the corresponding PL combination. \%Diff represents the percentage of difference between MedianMPLB and MedianSPLB, i.e.,
$\%Diff=(MedianMPLB-MedianSPLB)/MedianSPLB\times{100\%} .$
\#RMPLB and \#RSPLB denote the numbers of reopened bugs out of MPLBs and SPLBs, respectively. \#MPLB and \#SPLB denote the numbers of MPLBs and SPLBs, respectively. \%RMPLB and \%RSPLB denote the percentages of \#RMPLB over \#MPLB and \#RSPLB over \#SPLB, respectively. \textit{P-value} denotes the \textit{p-value} of Mann-Whitney U or Chi-squared test.

\textbf{Change Complexity:}
As shown in Table \ref{table-MPLBs-SPLBs}, the medians of the four resolution characteristics regarding change complexity for MPLBs involving the 13 PL combinations tend to be larger than that of the corresponding SPLBs. Specifically, LOCM for MPLBs is 17.07\%-623.86\% larger than that for SPLBs, NOFM for MPLBs is 50.00\%-450.00\% larger than that for SPLBs, NODM for MPLBs is 50.00\%-250.00\% larger than that for SPLBs, and Entropy for MPLBs is 1.31\%-82.94\% larger than that for SPLBs except for the PL combination of Clojure and Java. We further performed Mann-Whitney U tests on the four change complexity measures for MPLBs and SPLBs, and the \textit{p-value} for each measure is less than 0.05 for each PL combination (except \textit{p-value} of 0.267 for Entropy of the PL combination of Clojure and Java). It indicates that the change complexity of the resolution of MPLBs is significantly greater than that of SPLBs. 

\textbf{Open Time:} 
As presented in Table \ref{table-MPLBs-SPLBs}, the Mann-Whitney U test results reveal that the median OT of the MPLBs involving most (12 out 13, 92.3\%) PL combinations is not significantly shorter than that of the corresponding SPLBs.  
Specifically, there are no significant differences on OT between MPLBs involving 3 PL combinations (Java, Python; C\#, Java; JavaScript, Python) and that of the corresponding SPLBs; the median OT of the MPLBs involving one PL combination (Erlang, JavaScript) is significantly smaller than that of the corresponding SPLBs; and the median OT of MPLBs involving the rest 9 PL combinations is significantly (from 19.52\% to 529.57\%) longer than that of the corresponding SPLBs. 

\textbf{Reopen Rate:} The percentages of the reopened MPLBs and SPLBs are shown in TABLE \ref{table-MPLBs-SPLBs}, in which the percentages of reopened MPLBs and SPLBs range from 1.75\% to 20.66\% and from 4.89\% to 6.28\%, respectively. In addition, MPLBs involving 9 out of 13 PL combinations are not significantly associated with bug reopen, MPLBs involving one PL combination (C/C++, Python) is significantly negatively associated with bug reopen, and MPLBs involving 3 PL combinations (Java, Python; Java, JavaScript; JavaScript, Python) are significantly positively associated with bug reopen. 

\begin{table*}[]
\caption{ Indicators of six resolution characteristics of MPLBs with 13 PL combinations compared with corresponding SPLBs (RQ4). }
\scriptsize
\centering
\resizebox{\linewidth}{!}{
\begin{tabular}{|
>{\columncolor[HTML]{BDD7EE}}c |
>{\columncolor[HTML]{DDEBF7}}c |c|c|c|c|c|c|c|c|c|c|c|c|c|}
\hline
\multicolumn{2}{|c|}{\cellcolor[HTML]{BDD7EE}\begin{tabular}[c]{@{}c@{}}Bug resolution\\characteristic\end{tabular}} & \cellcolor[HTML]{DDEBF7}\begin{tabular}[c]{@{}c@{}}Java,\\ Scala\end{tabular} & \cellcolor[HTML]{DDEBF7}\begin{tabular}[c]{@{}c@{}}C/C++,\\ Python\end{tabular} & \cellcolor[HTML]{DDEBF7}\begin{tabular}[c]{@{}c@{}}Java,\\ Python\end{tabular} & \cellcolor[HTML]{DDEBF7}\begin{tabular}[c]{@{}c@{}}Java,\\ JavaScript\end{tabular} & \cellcolor[HTML]{DDEBF7}\begin{tabular}[c]{@{}c@{}}Python,\\ Scala\end{tabular} & \cellcolor[HTML]{DDEBF7}\begin{tabular}[c]{@{}c@{}}C/C++,\\ Java\end{tabular} & \cellcolor[HTML]{DDEBF7}\begin{tabular}[c]{@{}c@{}}C\#,\\ Java\end{tabular} & \cellcolor[HTML]{DDEBF7}\begin{tabular}[c]{@{}c@{}}JavaScript,\\ Python\end{tabular} & \cellcolor[HTML]{DDEBF7}\begin{tabular}[c]{@{}c@{}}C/C++,\\ Java,\\ Python\end{tabular} & \cellcolor[HTML]{DDEBF7}\begin{tabular}[c]{@{}c@{}}Erlang,\\ JavaScript\end{tabular} & \cellcolor[HTML]{DDEBF7}\begin{tabular}[c]{@{}c@{}}JavaScript,\\ PHP\end{tabular} & \cellcolor[HTML]{DDEBF7}\begin{tabular}[c]{@{}c@{}}Clojure,\\ Java\end{tabular} & \cellcolor[HTML]{DDEBF7}\begin{tabular}[c]{@{}c@{}}Java,\\ JavaScript,\\ Python\end{tabular} \\ \hline
\cellcolor[HTML]{BDD7EE}                           & MedianSPLB & 41                                                                            & 21                                                                              & 41                                                                             & 41                                                                                 & 33                                                                              & 41                                                                            & 44                                                                          & 19                                                                                   & 41                                                                                      & 17                                                                                   & 17                                                                                & 41                                                                              & 41                                                                                           \\ \cline{2-15} 
\cellcolor[HTML]{BDD7EE}                           & MedianMPLB     & 134.5                                                                         & 87                                                                              & 140                                                                            & 106                                                                                & 61.5                                                                            & 99                                                                            & 318.5                                                                       & 70                                                                                   & 288                                                                                     & 28                                                                                   & 68.5                                                                              & 48                                                                              & 258                                                                                          \\ \cline{2-15} 
\cellcolor[HTML]{BDD7EE}                           & \%Diff     & 228.05\%                                                                      & 314.29\%                                                                        & 241.46\%                                                                       & 158.54\%                                                                           & 86.36\%                                                                         & 141.46\%                                                                      & 623.86\%                                                                    & 268.42\%                                                                             & 602.44\%                                                                                & 64.71\%                                                                              & 302.94\%                                                                          & 17.07\%                                                                         & 529.27\%                                                                                     \\ \cline{2-15} 
\multirow{-4}{*}{\rotatebox{90}{\cellcolor[HTML]{BDD7EE}LOCM}}     & \textit{p-value}    & \textless 0.001                                                               & \textless 0.001                                                                 & \textless 0.001                                                                & \textless 0.001                                                                    & \textless 0.001                                                                 & \textless 0.001                                                               & \textless 0.001                                                             & \textless 0.001                                                                      & \textless 0.001                                                                         & \textless 0.001                                                                      & \textless 0.001                                                                   & 0.010                                                                 & \textless 0.001                                                                              \\ \hline
\cellcolor[HTML]{BDD7EE}                           & MedianSPLB & 2                                                                             & 2                                                                               & 2                                                                              & 2                                                                                  & 2                                                                               & 2                                                                             & 2                                                                           & 2                                                                                    & 2                                                                                       & 2                                                                                    & 2                                                                                 & 2                                                                               & 2                                                                                            \\ \cline{2-15} 
\cellcolor[HTML]{BDD7EE}                           & MedianMPLB     & 6                                                                             & 3                                                                               & 5                                                                              & 5                                                                                  & 4                                                                               & 5                                                                             & 8                                                                           & 4                                                                                    & 10.5                                                                                    & 3                                                                                    & 4                                                                                 & 3                                                                               & 11                                                                                           \\ \cline{2-15} 
\cellcolor[HTML]{BDD7EE}                           & \%Diff     & 200.00\%                                                                      & 50.00\%                                                                         & 150.00\%                                                                       & 150.00\%                                                                           & 100.00\%                                                                        & 150.00\%                                                                      & 300.00\%                                                                    & 100.00\%                                                                             & 425.00\%                                                                                & 50.00\%                                                                              & 100.00\%                                                                          & 50.00\%                                                                         & 450.00\%                                                                                     \\ \cline{2-15} 
\multirow{-4}{*}{\rotatebox{90}{\cellcolor[HTML]{BDD7EE}NOFM}}     & \textit{p-value}    & \textless 0.001                                                               & \textless 0.001                                                                 & \textless 0.001                                                                & \textless 0.001                                                                    & \textless 0.001                                                                 & \textless 0.001                                                               & \textless 0.001                                                             & \textless 0.001                                                                      & \textless 0.001                                                                         & \textless 0.001                                                                      & \textless 0.001                                                                   & \textless 0.001                                                                 & \textless 0.001                                                                              \\ \hline
\cellcolor[HTML]{BDD7EE}                           & MedianSPLB & 2                                                                             & 1                                                                               & 2                                                                              & 2                                                                                  & 2                                                                               & 2                                                                             & 2                                                                           & 1                                                                                    & 2                                                                                       & 1                                                                                    & 1                                                                                 & 2                                                                               & 2                                                                                            \\ \cline{2-15} 
\cellcolor[HTML]{BDD7EE}                           & MedianMPLB     & 4                                                                             & 2                                                                               & 4                                                                              & 4                                                                                  & 3                                                                               & 3                                                                             & 6                                                                           & 3                                                                                    & 6                                                                                       & 2                                                                                    & 2                                                                                 & 3                                                                               & 7                                                                                            \\ \cline{2-15} 
\cellcolor[HTML]{BDD7EE}                           & \%Diff     & 100.00\%                                                                      & 100.00\%                                                                        & 100.00\%                                                                       & 100.00\%                                                                           & 50.00\%                                                                         & 50.00\%                                                                       & 200.00\%                                                                    & 200.00\%                                                                             & 200.00\%                                                                                & 100.00\%                                                                             & 100.00\%                                                                          & 50.00\%                                                                         & 250.00\%                                                                                     \\ \cline{2-15} 
\multirow{-4}{*}{\rotatebox{90}{\cellcolor[HTML]{BDD7EE}NODM}}     & \textit{p-value}    & \textless 0.001                                                               & \textless 0.001                                                                 & \textless 0.001                                                                & \textless 0.001                                                                    & \textless 0.001                                                                 & \textless 0.001                                                               & \textless 0.001                                                             & \textless 0.001                                                                      & \textless 0.001                                                                         & \textless 0.001                                                                      & \textless 0.001                                                                   & \textless 0.001                                                                 & \textless 0.001                                                                              \\ \hline
\cellcolor[HTML]{BDD7EE}                           & MedianSPLB & 0.881  & 0.787   & 0.881  & 0.881  & 0.870  & 0.881  & 0.918  & 0.787   & 0.881  & 0.504   & 0.504   & 0.881   & 0.881  \\ \cline{2-15} 
\cellcolor[HTML]{BDD7EE}                           & MedianMPLB     & 0.927  & 0.918   & 0.918  & 0.918  & 0.896  & 0.932  & 0.930  & 0.910   & 0.936  & 0.892   & 0.922   & 0.873   & 0.885   \\ \cline{2-15} 
\cellcolor[HTML]{BDD7EE}                           & \%Diff     & 5.22\% & 16.65\% & 4.20\% & 4.20\% & 2.99\% & 5.79\% & 1.31\% & 15.63\% & 6.24\% & 76.98\% & 82.94\% & -0.91\% & 0.45\%     \\ \cline{2-15} 
\multirow{-4}{*}{\rotatebox{90}{\cellcolor[HTML]{BDD7EE}Entropy}}  & \textit{p-value}    & \textless 0.001                                                               & \textless 0.001                                                                 & \textless 0.001                                                                & \textless 0.001                                                                    & \textless 0.001                                                                 & \textless 0.001                                                               & \textless 0.001                                                             & \textless 0.001                                                                      & \textless 0.001                                                                         & \textless 0.001                                                                      & \textless 0.001                                                                   & 0.267                                                                           & 0.028                                                                               \\ \hline
\cellcolor[HTML]{BDD7EE}                           & MedianSPLB & 5.901                                                                         & 7.062                                                                           & 5.901                                                                          & 5.901                                                                              & 5.024                                                                           & 7.062                                                                         & 5.901                                                                       & 2.002                                                                                & 7.062                                                                                   & 7.967                                                                                & 6.211                                                                             & 7.600                                                                           & 5.901                                                                                        \\ \cline{2-15} 
\cellcolor[HTML]{BDD7EE}                           & MedianMPLB     & 8.985                                                                         & 22.070                                                                          & 4.844                                                                          & 7.053                                                                              & 7.381                                                                           & 19.630                                                                        & 7.256                                                                       & 1.903                                                                                & 44.460                                                                                  & 5.886                                                                                & 10.120                                                                            & 12.140                                                                          & 9.376                                                                                        \\ \cline{2-15} 
\cellcolor[HTML]{BDD7EE}                           & \%Diff     & 52.26\%                                                                       & 212.52\%                                                                        & -17.91\%                                               & 19.52\%                                                                            & 46.91\%                                                                         & 177.97\%                                                                      & 22.96\%                                                                     & \-4.95\%                                                      & 529.57\%                                                                                & -26.12\%                                                     & 62.94\%                                                                           & 59.74\%                                                                         & 58.89\%                                                                                      \\ \cline{2-15} 
\multirow{-4}{*}{\rotatebox{90}{\cellcolor[HTML]{BDD7EE}OT}}       & \textit{p-value}    & \textless 0.001                                                               & \textless 0.001                                                                 & 0.607                                                                          & \textless 0.001                                                                    & \textless 0.001                                                                 & \textless 0.001                                                               & 0.084                                                                       & 0.111                                                                                & \textless 0.001                                                                         & \textless 0.001                                                                      & \textless 0.001                                                                   & 0.003                                                                  & 0.025                                                                              \\ \hline
\cellcolor[HTML]{BDD7EE}                           & \#RSPLB   & 2032                                                                          & 908                                                                             & 2024                                                                           & 2031                                                                               & 708                                                                             & 2232                                                                          & 1695                                                                        & 707                                                                                  & 2582                                                                                    & 370                                                                                  & 368                                                                               & 1683                                                                            & 2381                                                                                         \\ \cline{2-15} 
\cellcolor[HTML]{BDD7EE}                           & \#SPLB    & 36684  & 14467           & 35711  & 36073           & 14471  & 36680  & 29439  & 13860           & 43429  & 7234   & 7523   & 29053  & 42822                                                                                       \\ \cline{2-15} 
\cellcolor[HTML]{BDD7EE}                           & \%RSPLB   & 5.54\% & 6.28\%          & 5.67\% & 5.63\%          & 4.89\% & 6.09\% & 5.76\% & 5.10\%          & 5.95\% & 5.11\% & 4.89\% & 5.79\% & 5.56\%  \\ \cline{2-15} 
\cellcolor[HTML]{BDD7EE}                           & \#RMPLB   & 101                                                                           & 65                                                                              & 71                                                                             & 85                                                                                 & 16                                                                              & 9                                                                             & 9                                                                           & 25                                                                                   & 6                                                                                       & 6                                                                                    & 2                                                                                 & 1                                                                               & 4                                                                                            \\ \cline{2-15} 
\cellcolor[HTML]{BDD7EE}                           & \#MPLB    & 1782   & 1635            & 903    & 799             & 266    & 246    & 122    & 121             & 86     & 63     & 62     & 57     & 52        \\ \cline{2-15} 
\cellcolor[HTML]{BDD7EE}                           & \%RMPLB  & 5.67\% & 3.98\%          & 7.86\% & 10.64\%         & 6.02\% & 3.66\% & 7.38\% & 20.66\%         & 6.98\% & 9.52\% & 3.23\% & 1.75\% & 7.69\%     \\ \cline{2-15} 
\multirow{-7}{*}{\rotatebox{90}{\cellcolor[HTML]{BDD7EE}Reopen}} & \textit{p-value}    & 0.817                                                                         & \textless 0.001                                                                 & 0.005                                                                          & \textless 0.001                                                                    & 0.401                                                                           & 0.112                                                                         & 0.444                                                                       & \textless 0.001                                                                      & 0.686                                                                                   & 0.197                                                                                & 0.756                                                                             & 0.307                                                                           & 0.713                                                                                        \\ \hline
\end{tabular}
}
\label{table-MPLBs-SPLBs}
\end{table*}



\subsection{Causes for Involving Multiple PLs in Bug Resolution (RQ5)}

To understand the causes why the resolution of MPLBs involves multiple PLs, we manually analyzed a sample set of the MPLBs (i.e., 931 MPLBs), including all 28 MPLBs involving more than 3 PLs, all 302 MPLBs involving 3 PLs, and a sampled subset of 601 MPLBs out of all 6,370 MPLBs involving 2 PLs.
After two rounds of pilot bug labelling, the Fleiss Kappa value for the labelling results of the second and third authors was 0.76, greater than the threshold 0.75.

\subsubsection{Bug Resolution Categories}
\begin{figure}
\centerline{\includegraphics[width=1.3in]{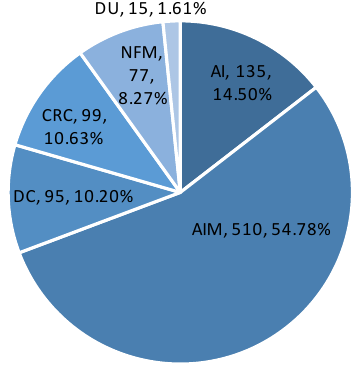}}
\caption{Number and percentage of the six causes for involving multiple PLs in bug resolution (RQ5).}
\label{fig-RQ5ClassData}
\end{figure}

After several rounds of manual analysis, we identified the following 6 categories of bug resolution involving multiple PLs, as shown in Fig. \ref{fig-RQ5ClassData}:

\textbf{(1) Algorithm Implementation (AI).} 
This category of bug resolution solves an MPLB by new algorithm (function) or code logic implementation. For example, MPLB ``AMBARI-14948: Config consistency checker'' is resolved by implementing the \textsc{check\_database()} function in Java and Python.

\textbf{(2) Algorithm Implementation Modifications (AIM).} 
This category of bug resolution solves an MPLB by modifying existing algorithm implementation, function call, etc. For instance, MPLB ``SPARK-19134: Fix several sql, mllib and status api examples not working'' is resolved by modifying algorithm implementation by Java, Python, and Scala.


\textbf{(3) Data-related Changes (DC).} 
This category of bug resolution indicates an MPLB resolved through data-related changes, including but not limited to: time format, data type, type precision, and data structure changes (adding variables to the data structure). For example, MPLB ``DISPATCH-284: Added a connection id to link which can be used to linked back to identity of connection" is fixed by adding a \textsc{uint64\_t management\_id} variable in C and Python. 

\textbf{(4) Configuration-related Changes (CRC).} 
This category of bug resolution indicates that an MPLB is solved via configuration-related changes to the software system, such as making configuration option (file)-related changes, fixing dependency issues between components, compatibility issues, and hardcoded issues. For example, MPLB ``AMBARI-6484: Hbase RegionServer -Xmn must be configurable" is resolved by adjusting the \textsc{hbase\_regionserver\_xmn} parameter from hardcoded to adjustable in JavaScript and Python.


\textbf{(5) Non-functional Modifications (NFM).} 
This category of bug resolution fixes an MPLB by non-functional modifications, such as specification of function/variable names, removal of dead code, optimization of code. For example, MPLB ``SPARK-2739: Rename registerAsTable to registerTempTable'' was assigned a Jira priority of ``Blocker''; users complained that it was difficult to differentiate between \textsc{registerAsTable} and \textsc{saveAsTable}, thus \textsc{registerAsTable} function name was changed to \textsc{registerTempTable} in Java, Python, and Scala.


\textbf{(6) Documentation Updates (DU).} 
This category of bug resolution solves an MPLB by updating documents, such as annotated documents or examples. For instance, MPLB ``SPARK-32035: Inconsistent AWS environment variables in documentation" is fixed through changing annotations, i.e., changing annotation \textsc{AWS\_SECRET\_KEY} to \textsc{AWS\_SECRET\_ACCESS\_KEY}. This annotation change involves source files written in Java, Python, and Scala.


\subsubsection{Cross-language Calling Mechanisms}
\begin{figure}
\centerline{\includegraphics[width=3.3in]{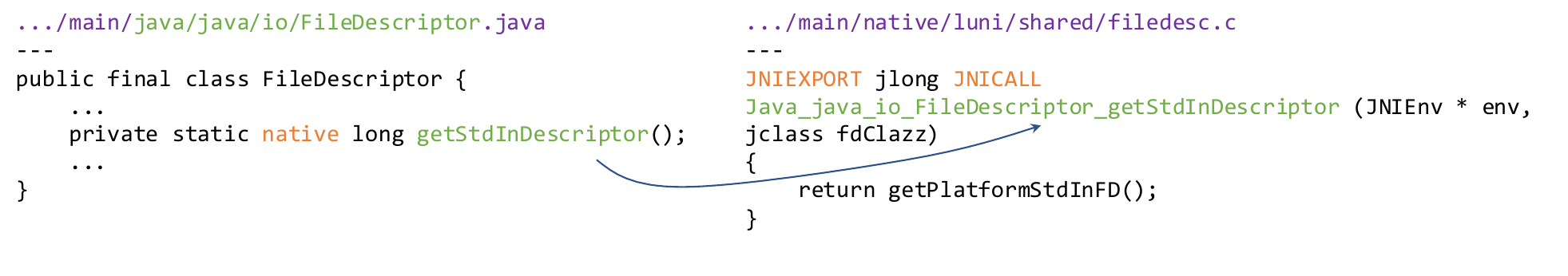}}
\caption{JNI native library mechanism for cross-language calls (RQ5).}
\label{fig-JNI}
\end{figure}

\begin{figure}
\centerline{\includegraphics[width=3.3in]{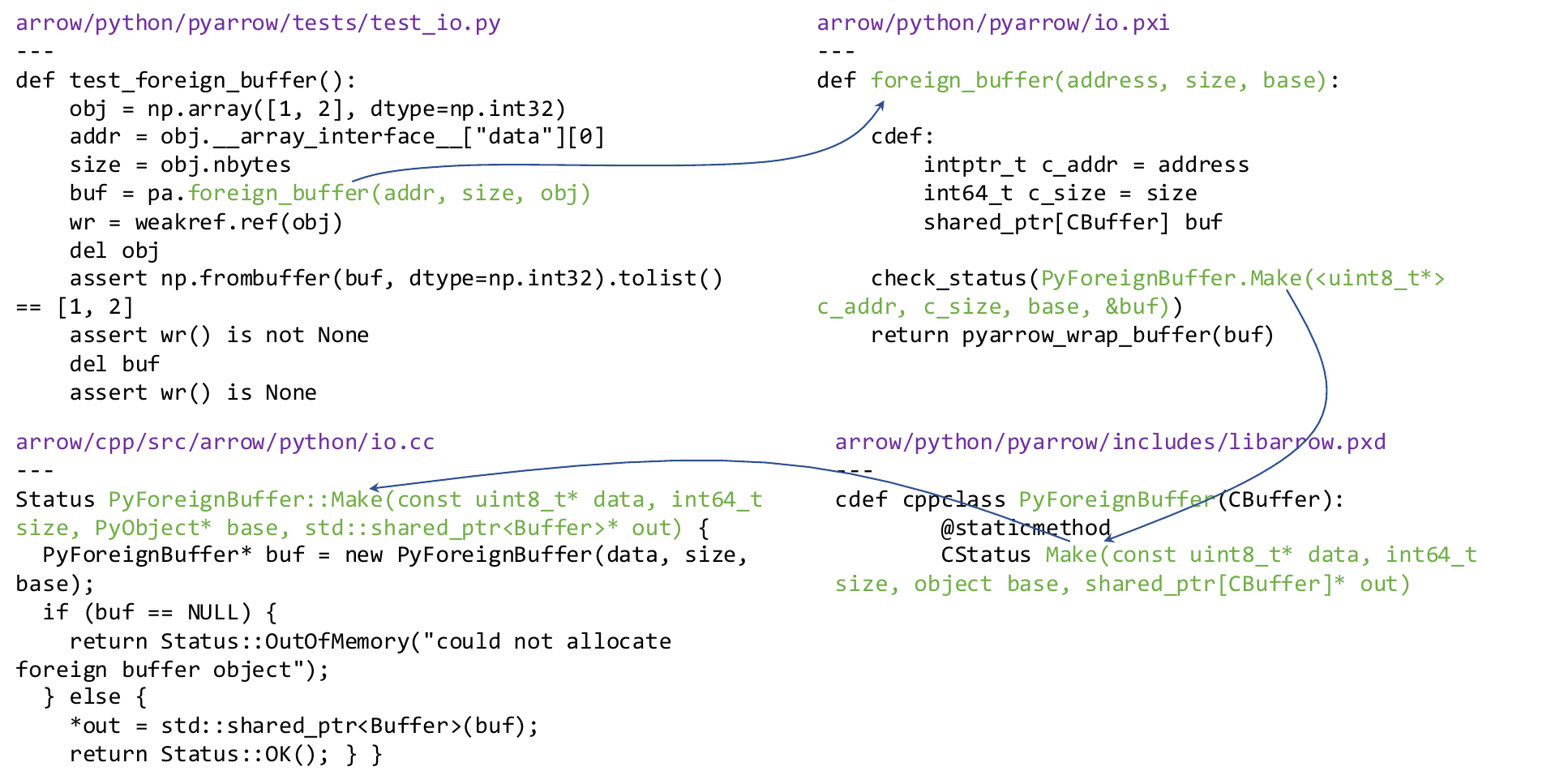}}
\caption{Cython native library mechanism for cross-language calls (RQ5).}
\label{fig-Cython}
\end{figure}

We looked further into cross-language calls to explain the causes for involving multiple PLs in bug resolution, and found 6 cross-language calling mechanisms used by the involved source files written in multiple PLs. These mechanisms are described as follows: 

\textbf{(1) Local Library Mechanisms (LLM).}  Through an LLM, code in one PL can directly call another PL's native libraries, e.g., Java Native Interface (JNI) and Dynamic-link library (DLL). We briefly introduce two LLMs in the following.
\textbf{(a) JNI} is a native programming interface and it allows Java code that runs inside a Java Virtual Machine (JVM) to interoperate with applications and libraries written in other PLs, such as C, C++, and assembly \cite{JNIdoc}. For example, MPLB ``HARMONY-6642: [classlib][luni] FileInputStream doesn't close FD in native code'' is resolved by implementing native library calls via JNI methods, as shown in Fig. \ref{fig-JNI}. 
\textbf{(b) Cython} is a PL that makes writing C extensions for the Python language as easy as Python itself \cite{Cythondoc}. Cython 
code is translated into optimized C/C++ code (essentially DLLs) and compiled as Python extension modules. For example, to solve MPLB ``ARROW-2270: [Python] ForeignBuffer doesn't tie Python object lifetime to C++ buffer lifetime'', as shown in Fig.~\ref{fig-Cython}, Foreign Buffer is defined and implemented in C++ and wrapped in Cython, which enables Python to call the Foreign Buffer across PLs. 

\begin{figure}
\centerline{\includegraphics[width=3.2in]{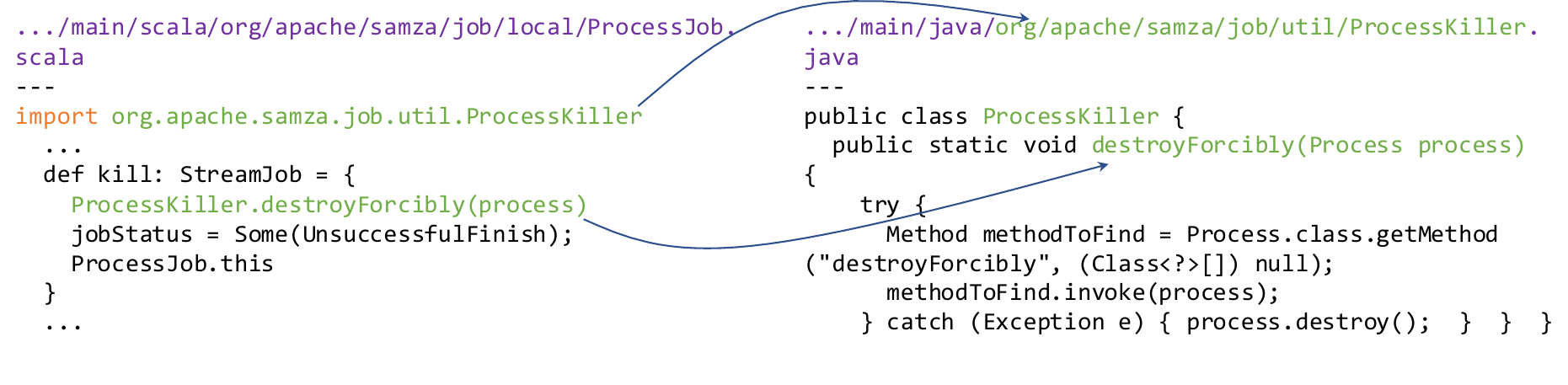}}
\caption{Common runtime mechanism for cross-language calls (RQ5).}
\label{fig-JVM}
\end{figure}

\textbf{(2) Common Runtime Mechanisms (CRM).}
Converting code in various PLs into intermediate code (e.g., bytecode for JVM or intermediate language code for .NET) that enables the code to run on the same runtime platform, allowing cross-language calls. 
For example, as shown in Fig. \ref{fig-JVM}, in the resolution of MPLB ``SAMZA-940: TestProcessJob.testProcessJobKillShouldWork fails occasionally'', Scala code can call Java code since both can run on JVM.

\begin{figure}
\centerline{\includegraphics[width=3.3in]{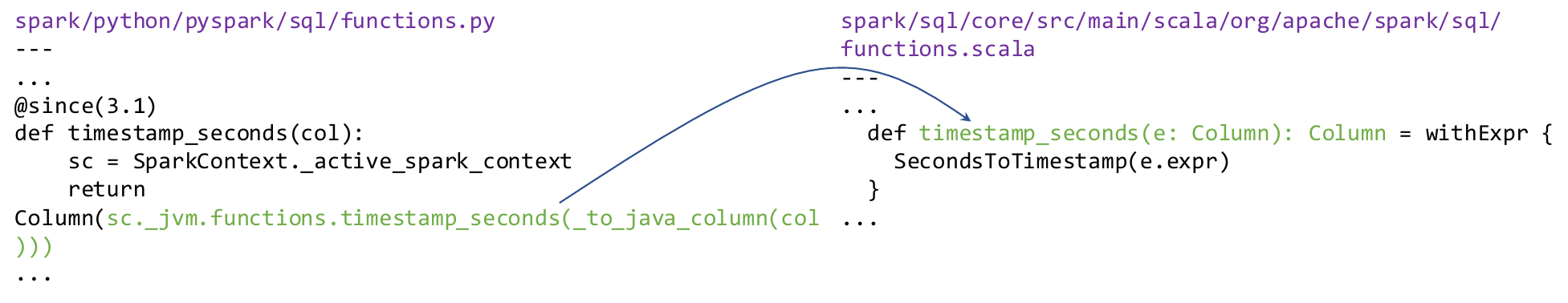}}
\caption{Py4J communication protocol mechanism for cross-language calls.}
\label{fig-Py4J}
\end{figure}

\begin{figure}
\centerline{\includegraphics[width=3.3in]{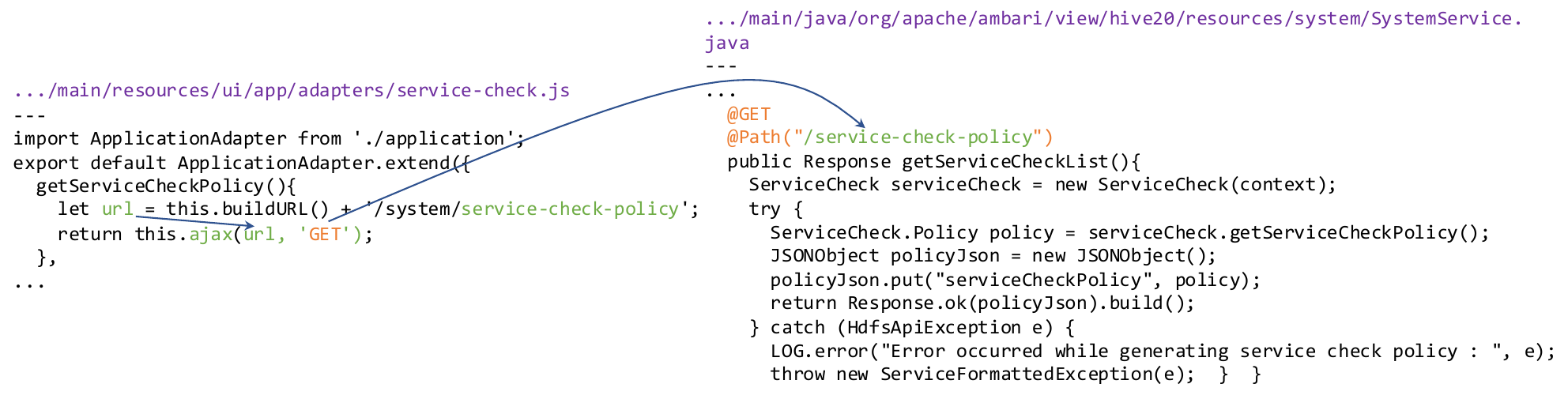}}
\caption{HTTP for cross-language calls (RQ5).}
\label{fig-HTTP}
\end{figure}

\textbf{(3) Communication Protocol Mechanisms (CPM).} Cross-language calls between different PLs are accomplished through specific communication protocols, e.g., HTTP. We briefly introduce three CPMs in the following.
\textbf{(a) Thrift} is a cross-language service framework, which can automatically generate code in different PLs based on the service interface and data types defined in Interface Definition Language (IDL) \cite{Thriftdoc}. Thrift enables clients and servers written in different PLs to communicate through Remote Procedure Call (RPC). For example, in the resolution of MPLB ``AIRAVATA-1281: experiments returned by searchExperimentByName don't have applicationId'', \textsc{7: optional string applicationId} is added to the Thrift IDL file, and Thrift automatically generates code for C++, PHP, and Java, allowing for cross-language calling via RPC. 
\textbf{(b) Py4J} is an inter-process communication (IPC) technology widely used in Spark's pyspark library \cite{Py4Jdoc}. JVM acts as the server side of the IPC, starting a socket port to provide the service, and Python code acts as the client side of the IPC, calling the client interface provided by Py4J. For example, in the resolution of MPLB ``SPARK-31710: Fail casting numeric to timestamp by default", as shown in Fig. \ref{fig-Py4J}, the method \textsc{timestamp\_seconds (e:Column): Column} is defined in Scala, and Python code utilizes Py4J to enable cross-language calls.  
\textbf{(c) HTTP} refers to cross-language calls between different PLs through the HTTP protocol. The client in a PL sends an HTTP \textsc{GET/POST} request to the server in a distinct PL, which processes the request and returns a response. This is done using a wrapped HTTP library, such as Python's Request library \cite{Requestsdoc}. For example, in the resolution of MPLB ``AMBARI-20517: make home directory check as optional in hive20 view'', as shown in Fig. \ref{fig-HTTP}, JavaScript is used to implement the new \textsc{getServiceCheckPolicy()} method, which uses \textsc{GET} requests to make a cross-language call to the \textsc{service-check-policy} interface in Java.

\textbf{(4) Inter-language Testing (ILT). }
The cause for involving multiple PLs in inter-language testing is that source files in one or more PLs (e.g., C) are modified, 
while test cases written in other PLs (e.g., Python) for black-box testing need to be modified accordingly. 
For example, in the resolution of MPLB ``SVN-1581: svn cp url onto missing file corrupts wc'', the C source file now includes detection of logical obstructions, while the Python source file incorporates test cases for the corresponding black-box tests.


\textbf{(5) MPL Definition and Implementation (MDI).}
Source files in multiple PLs are modified for the same reason, since there is a link between MPL modifications, such as implementation or modification of the same algorithm in multiple PLs, unified definition or modification of data structures in multiple PLs. For example, in the resolution of MPLB ``THRIFT-3276: Binary data does not decode correctly using the TJSONProtocol when the base64 encoded data is padded'', the \textsc{ignore padding} algorithm was implemented in three different PLs, i.e., C++, C\#, and Java.


Due to our limited knowledge, experience, and resources in comprehending the source code of the 54 non-trivial projects across a wide range of application domains, we only identified 515 cross-language calls in the resolution of 505 MPLBs, which account for 54.2\% of the 931 MPLBs in total under manual analysis.
The distribution of MPLBs of each bug resolution category over the 5 cross-language calling mechanisms is shown in TABLE \ref{table-RQ5-Root}. 
We found that the MDI mechanism was used most frequently, with a total of 195 times. The ILT mechanism is the most commonly used mechanism in the AIM category, accounting for 37.8\% (107/283) of all the usages.

\begin{table}[]
\caption{Distribution of the number of the five cross-language calling mechanisms in AI, AIM, DC, CRC, NFM, and DU (RQ5).}
\scriptsize
\centering
\begin{tabular}{cccccccc}
\toprule
     & AI & AIM & DC & CRC & NFM & DU & Total \\ \midrule
LLM    & 10 & 16  & 3  & 2   & 0  & 0  & 31      \\
CRM    & 10 & 42  & 6  & 3   & 0  & 0  & 61      \\
CPM    & 24 & 51  & 10 & 5   & 1  & 0  & 91      \\
ILT    & 13 & 107 & 13 & 4   & 0  & 0  & 137     \\
MDI    & 21 & 67  & 27 & 26  & 39 & 15 & 195     \\ 
Total  & 78 & 283 & 59 & 40  & 40 & 15 & 515      \\ \bottomrule
\end{tabular}
\label{table-RQ5-Root}
\end{table}

\section{Discussion}
\label{chap:discus}

\subsection{Interpretation of Study Results}

\emph{\textbf{RQ1}}: 
Taking all projects as a whole, only about 10\% of the bugs are MPLBs, which indicates that a majority of bugs in MPL software development can be solved by modifying source code in a single PL. It further implies that inter-language dependencies tend to be well designed and implemented so that the impact of most bugs does not propagate across PLs.

\emph{\textbf{RQ2}}: The vast majority (95\%) of MPLBs involve source files written in only 2 PLs. A possible reason is that modifying source files involving more PLs in a single bug resolution may result in higher code change complexity, which requires a more comprehensive consideration and thus more effort when performing change impact analysis on the software system.

\emph{\textbf{RQ3}}: (1) The resolution characteristics of MPLBs show a more complex resolution process than SPLBs. This means that resolving MPLBs will increase the difficulty of software development. MPLBs involve source files written in multiple PLs, and the source files in different PLs usually distribute in different components; therefore, MPLBs tend to have a global impact on the software system, and the change complexity of MPLBs will be higher that of SPLBs.
(2) Bug resolution involving multiple PLs is related to the reopen of bugs, resulting in increased rework.

\emph{\textbf{RQ4}}: 
(1) Higher code change complexity of resolution of MPLBs does not necessarily mean longer OT of the MPLBs. For instance, among the MPLBs involving two PLs, MPLBs involving Java tend to have greater LOCM, NOFM, and NODM but take shorter OT; in contrast, MPLBs involving C/C++ tend to have smaller LOCM, NOFM, and NODM but take longer OT. 
This observation is similar to the conclusion on SPLBs reached by Zhang et al. \cite{zhang2019study}. 
(2) MPLBs involving C/C++, Java, and Python show a much larger OT than MPLBs involving other PL combinations. This may be related to almost the highest code change complexity of the resolution of those MPLBs. 
(3) The percentage of reopened MPLBs involving different PL combinations fluctuates strongly while the percentage of reopened SPLBs involving different PL combinations is relatively stable. This is partially because the numbers of MPLBs and reopened MPLBs involving different PL combinations are too small, compared with the numbers of SPLBs and reopened SPLBs involving the corresponding PL combinations.



\emph{\textbf{RQ5}}: (1) The classification of causes of bug resolution involving multiple PLs shows that involving multiple PLs in the resolution of 54.78\% of the MPLBs result from modifications to algorithm implementation (i.e., AIM). Inter-language testing (ILT) is used by the modified source files for resolving 107 out of 283 MPLBs by AIM, the largest proportion. This can be explained by that changes in algorithm implementation are often followed by corresponding testing. (2) MPL definitions and implementations (MDI) is used in the resolution of 195 out of 505 MPLBs, accounting for the largest proportion; the possible reason is that common implementations or modifications of multiple PLs aim to provide interfaces to multiple PLs to meet the calls from different PLs.

\subsection{Implications for Practitioners}

\textbf{Practitioners should be cautious of bug resolution involving the PL combination of JavaScript and Python.}
MPLBs involving the PL combination of JavaScript and Python have a reopen rate of 20.66\%, which is much higher than that of MPLBs involving other PL combinations. A reopened bug can take considerably more time and effort to fix \cite{tagra2022revisiting}.


\textbf{The code change complexity and OT are related to specific PLs.} For instance, MPLBs involving PL combinations with Java show a higher code change complexity and shorter OT while MPLBs involving PL combinations with C/C++ show a lower code change complexity and longer OT.

\textbf{Reopen of MPLBs is in association with specific PL combinations.} There are significantly positive associations between bug reopen and modifying source files in 3 PL combinations (Java, Python; Java, JavaScript; JavaScript, Python).

\textbf{The complexity of resolution of MPLBs involving most PL combinations are manageable.} This is evidenced by the fact that there is no significant association between bug reopen and modifying source files in 9 out 13 PL combinations.

\textbf{The identified causes for involving multiple PLs in bug resolution and cross-language calling mechanisms can help practitioners to better understand the links between MPL source code.} There has been limited research on the classification of cross-language calling mechanisms.

\subsection{Implications for Researchers}

\textbf{More PL combinations need to be explored with respect to the resolution characteristics of MPLBs.} 
It is valuable to get informed the impact of different PL combinations on the bug resolution, so that developers can make appropriate decisions on release planning and task assignment when facing MPLBs involving different PL combinations.


\textbf{There is a need of a more complete picture of cross-language calling mechanisms.} Due to the core role of cross-language calling mechanisms in the research on MPL topics, such a complete picture will save researchers considerable effort in MPL analysis. 


\section{Threats to Validity}\label{chap:threats}



\textbf{Construct validity} is concerned with whether the data we collected (listed in Table \ref{table:DataItemsForBug}) is consistent with the true values we expect. A possible threat to construct validity is that not all resolved bugs are associated with the corresponding commits. Due to various reasons, committers may not explicitly mention the resolved bug ID in the corresponding commit message, which may negatively influence the representativeness of the collected bugs and further affect the accuracy of entropy and bug OT. Through our manual check, we confirmed that the bugs with explicit links to corresponding commits are not reported in a narrow time span and also not resolved by a small group of specific developers. Therefore, this threat is partially alleviated.
Another potential threat is biases held by  different researchers during manual analysis, which may result in incorrect classification of MPLB resolution. To mitigate this threat, we adopted a three-step bug classification process, in which a pilot bug labelling was adopted to identify and resolve possible disagreements between researchers.


\textbf{External validity} centers around the generalizability of the study results. First, a possible threat to external validity is whether the selected projects are sufficiently representative. As described in Section \ref{CaseSelection}, we used a set of criteria to select the projects. All Apache projects satisfying the selection criteria were included. Thus, this threat was eliminated. Second, another threat was that only Apache MPL OSS projects were selected. This means that we cannot claim the validity of the study results for MPL projects from other OSS ecosystems.


\textbf{Internal validity} is the extent to which a piece of evidence supports a claim about cause and effect. In the data analysis for answering RQ5, we explored the causes why bug resolution involves multiple PLs, and selected a sample of 931 MPLBs out of the total 6700 MPLBs for our qualitative analysis. A threat is that other bugs that were not included in our studied sample may contain causes beyond what we identified in our analysis. However, since we selected the statistically representative sample with a 99\% confidence level and a 5\% margin of error, this sample selection bias could be alleviated.


\textbf{Reliability} refers to whether the same results can be produced when other researchers replicate this study. One potential threat is related to the implementation of the associated software tools used for data collection. These tools were primarily implemented by the second author, and the code for key functions was regularly reviewed by the first author. Another threat is related to the dataset used in this study. To increase the reliability of the study results, we provided the used dataset and classification results of manual analysis of bug resolution and cross-language calling mechanisms online \cite{Dataset}. Consequently, the threat to reliability was reduced.

\section{Conclusions and Future Work}
\label{conclusions}

This work aims to understand the resolution of MPLBs in MPL software systems and the reasons why bug resolution involves multiple PLs.
To this end, we conducted a large-scale empirical study on 66932 bugs with 6700 MPLBs included, coming from 54 MPL projects that were selected from 655 Apache OSS projects according to a set of inclusion criteria.  
The \textbf{main findings} are summarized as follows:
\textbf{(1)} The percentage of MPLBs in the selected projects ranges from 0.17\% to 42.26\%, and the percentage of MPLBs for all projects as a whole was 10.01\%.
\textbf{(2)} 95.0\% and 4.5\% of all the MPLBs involve source files in 2 and 3 PLs, respectively; this means that the vast majority of MPLBs involve source files written in only 2 PLs, while it is uncommon to modify source files in 4 or more PLs in MPLB resolution.
\textbf{(3)} The change complexity resolution characteristics and open time of MPLBs tend to be higher than those of SPLBs.
\textbf{(4)} There is significant association between MPLBs and bug reopen for specific PL combinations. The reopen rate of the combination of JavaScript and Python reaches 20.66\%.
\textbf{(5)} In the identified six causes for involving multiple PLs in bug resolution, Algorithm Implementation and Modification (AIM) accounts for the highest percentage (54.78\%), and five cross-language calling mechanisms were identified to deepen the understanding on MPLB resolution in MPL software systems.

We plan to further: (1) investigate the impact of MPLBs on MPL software architectures, e.g., whether MPLBs are relevant to the introduction of architectural technical debt, and (2) explore how to automatically identify the cross-language calling mechanisms manually identified in this study.

\balance

\bibliographystyle{IEEEtran}
\bibliography{references}
\end{document}